\documentclass[showpacs,aps,floatfix]{revtex4}
\usepackage{graphicx,bm,color}
\usepackage{subfigure}
\usepackage{amssymb}
\usepackage{amsmath}
\begin{document}
\title{Correlation between conserved charges in PNJL Model with multi-quark interactions} 

\author{Abhijit Bhattacharyya}
\email{abphy@caluniv.ac.in}
\author{Paramita Deb}
\email{paramita.deb83@gmail.com}
\affiliation{Department of Physics, University of Calcutta,
92, A. P. C. Road, Kolkata - 700009, INDIA}
\author{Anirban Lahiri}
\email{anirbanlahiri.boseinst@gmail.com}
\author{Rajarshi Ray}
\email{rajarshi@bosemain.boseinst.ac.in}
\affiliation{Center for Astroparticle Physics \&
Space Science, Bose Institute. Block-EN, Sector-V, Salt Lake, Kolkata-700091, INDIA 
 \\ \& \\ 
Department of Physics, Bose Institute, \\
93/1, A. P. C Road, Kolkata - 700009, INDIA}

\vskip 0.3in
\begin {abstract}
We present a study of correlations among conserved charges like baryon number, 
electric charge and strangeness in the framework of 2+1 flavor Polyakov 
loop extended Nambu-Jona-Lasinio model at vanishing chemical potentials,
up to fourth order. Correlations up to second order have been measured 
in Lattice QCD which compares well with our estimates given the inherent 
difference in the pion masses in the two systems. Possible physical 
implications of these correlations and their importance in understanding
the matter obtained in heavy-ion collisions are discussed. We also 
present comparison of the results with the commonly used unbound 
effective potential in the quark sector of this model. 
\end{abstract}
\maketitle

\vskip 0.3in
{\section{Introduction}
 It is now well established that strongly interacting matter may
exhibit a variety of phases depending on the ambient thermodynamic
conditions. We are on the way to draw the phase diagram of 
quantum chromodynamics (QCD) which is the theory of strong interactions.
The difficulty that we still encounter is to work with relatively
strong coupling strengths in the theory. The best way to go about
is to perform numerical simulation on the discretized version of 
QCD - the so called Lattice QCD (LQCD). This formulation however has
not got rid of all its inherent technical problems and it would take 
some time to find the final answers 
\cite{boyd,engels,fodor1,fodor2,allton1,allton2,allton3,forcrand,
aoki1,aoki2,arriola1,arriola2,arriola3,arriola4}.
Meanwhile one can look into the properties of strongly interacting
matter through effective models of QCD. Polyakov loop extended
Nambu-Jona-Lasinio (PNJL) model is one such model that
successfully captures various properties of strongly interacting matter
\cite{fuku1,ratti1,pisarski,fuku2,ratti2,gatto,ghosh,rayvdm,
deb1,osipov1,osipov2,osipov3,osipov4,kashiwa1,kashiwa2,deb2}.
The question as to exactly what extent can this model emulate
QCD is still a matter under investigation. The best way to 
judge it is to measure several sensitive quantities in this model
and contrast some of them to that available in the LQCD measurements.
The correlations among conserved charges are some such quantities 
that we intend to investigate here. Essentially, the fluctuations and 
correlations of conserved charges and their higher order cumulants 
provide information about the degrees of freedom of strongly 
interacting matter. These can be extracted from the PNJL model 
and LQCD through the study of diagonal and off-diagonal susceptibilities 
respectively. They can also provide information about the
existence of critical behavior, if any. This phenomenological
study is important as there is a rapid progress in the experimental
front at the facilities at CERN and Brookhaven heavy-ion colliders.
\par
On the Lattice, some of these correlators have been measured
at zero chemical potential. With 2 flavors it was shown 
\cite{gottlieb,gavai,bernard1, bernard2} that the fluctuations rise 
rapidly around the crossover region from hadronic matter to quasi-free 
quark matter. The higher order cumulants show 
non-monotonic behavior \cite{allton3,ejiri}. Similar measurements have 
been made in 2 flavor PNJL model with three-momentum cutoff
regularization \cite{ghosh,roessner1,friman,ray2}.
Quark number susceptibility (QNS) at finite density 
has been estimated in some works within 2 flavor
PNJL model \cite{roessner2}. Also, QNS has been studied in Hard Thermal Loop
approximation \cite{mustafa1,mustafa2,mustafa3}. Recently the idea of the 
numerical Taylor
expansion in terms of chemical potential for PNJL model has been 
used within the constraint that the net strange quark density is zero,
which is the case in ultra relativistic heavy ion collision \cite{fuku3}.
\par
For 2+1 flavors, fluctuations have been recently measured in LQCD 
\cite{cheng1,cheng,fodor10} as well as in PNJL model both with 
the usual unbound effective potential (UEP) \cite{fuku2,wu1,wu2} and 
with the bound effective potential (BEP) \cite{deb3}. Similar
calculations have been carried out in Polyakov loop coupled quark-meson
(PQM) model \cite{schaefer1,schaefer2,schaefer3,schaefer4} and 
its renormalization group improved version \cite{skokov}. 
\par
In this work we investigate the off-diagonal susceptibilities
which give the correlations among different conserved charges. 
Our paper is organized as follows. In Sec. \ref{formlsm}, we discuss 
the basic formalism of the PNJL model as well as the method of 
extracting the Taylor expansion coefficients of pressure that gives
the various susceptibilities. In section \ref{results} we present and 
discuss our results together with a comparison with the data obtained 
in LQCD. The last section contains a summary and our conclusions.
 
\section{Formalism}
\label{formlsm}}

{\subsection {Thermodynamic Potential}}
  2+1 flavor PNJL model with unbound effective potential has been 
studied elaborately in a number of recent works 
\cite{ratti1,fuku1,gatto,deb1}. To introduce a bound in the 
effective potential eight quark interaction terms have been 
introduced in 2+1 flavor NJL model \cite{osipov1,osipov2,osipov3,osipov4} 
and in the 2 flavor PNJL model \cite{kashiwa1,kashiwa2}. 
We developed the 2+1 flavor PNJL model with bound effective potential 
to study finite temperature and chemical potential properties \cite{deb2} 
within three-momentum cutoff regularization scheme. Comparing with available
LQCD data the model was shown to reproduce various aspects of QCD 
thermodynamics quite satisfactorily. We shall be using this model in the
present work. The relevant thermodynamic potential in the mean field 
approximation can be written as \cite{deb2}, 
\begin {align}
 \Omega &= {\cal {U^\prime}}[\Phi,\bar \Phi,T]+2{g_S}{\sum_{f=u,d,s}}
{\sigma_f^2}-\frac{g_D}{2}{\sigma_u}
{\sigma_d}{\sigma_s}+3\frac{g_1}{2}({\sum_{f=u,d,s}}{\sigma_f}^2)^2\nonumber\\
&+3{g_2}{\sum_{f=u,d,s}}{\sigma_f^4}-6{\sum_{f=u,d,s}}{\int_{0}^{\Lambda}}
     {\frac{d^3p}{{(2\pi)}^3}} E_{f}\Theta {(\Lambda-{ |\vec p|})}\nonumber \\
       &-2T{\sum_{f=u,d,s}}{\int_0^\infty}{\frac{d^3p}{{(2\pi)}^3}}
       \ln\left[1+3(\Phi+{\bar \Phi}e^{-\frac{(E_{f}-\mu_f)}{T}})
       e^{-\frac{(E_{f}-\mu_f)}{T}}+e^{-\frac{3(E_{f}-\mu_f)}{T}}\right]\nonumber\\
       &-2T{\sum_{f=u,d,s}}{\int_0^\infty}{\frac{d^3p}{{(2\pi)}^3}}
        \ln\left[1+3({\bar \Phi}+{ \Phi}e^{-\frac{(E_{f}+\mu_f)}{T}})
       e^{-\frac{(E_{f}+\mu_f)}{T}}+e^{-\frac{3(E_{f}+\mu_f)}{T}}\right]
\end {align}
where $g_S$ and $g_D$ are the four quark and six quark coupling constant and 
$g_1$ and $g_2$ are the eight quark
coupling constant. Here $\sigma_f=\langle{\bar \psi_f} \psi_f\rangle$
denotes chiral condensate of the quark with flavor $f$ and
$E_{f}=\sqrt {p^2+M^2_f}$ is the single quasi-particle energy. 
Here, constituent mass $M_f$ of flavor $f$ is given by the self-consistent
gap equation;
\begin{equation*}
M_f=m_f-2g_S\sigma_f+{\frac{g_D}{2}}\sigma_{f+1}\sigma_{f+2}-2g_1\sigma_f
(\sigma_u^2+\sigma_d^2+\sigma_s^2)-4g_2\sigma_f^3
\end{equation*}
where $f$, $f+1$ and $f+2$ take the labels of flavor $u$, $d$ and $s$
in cyclic order. In the above expression, the vacuum part integral has a 
ultraviolet cutoff $\Lambda$. For fixing the parameters 
$m_s$, $\Lambda$, $g_S$, $g_D$, $g_1$, $g_2$ we have used the following 
physical conditions \cite{deb2};
\begin{eqnarray*}
m_\pi=138 ~ {\rm MeV} ~~~ m_K &=& 494 ~ {\rm MeV} ~~~ 
m_\eta=480 ~ {\rm MeV} ~~~ m_{\eta\prime}=957 ~ {\rm MeV} \\
f_\pi &=& 93 ~ {\rm MeV} ~~~ f_K=117 ~ {\rm MeV} 
\end{eqnarray*}
and $m_u$ is kept fixed at 5.5 MeV. The parameters are given in 
table \ref{table1} for UEP and BEP.

The Polyakov loop $\Phi$ and its charge conjugate $\bar \Phi$ are defined as, 
\begin {equation*}
\Phi = (\rm{Tr}_c {\bf L})/N_c, {\hspace{0.3in}} {\bar \Phi} 
     = (\rm{Tr}_c {\bf L}^\dagger)/N_c
\end {equation*}
where, {\bf L} is the Wilson line given by,
\begin{equation*}
 {\bf L}=\left[\mathcal{P}\exp\left(i\int_{0}^{\beta}
A_4d\tau\right)\right]=\exp\left[\frac{iA_4}{T}\right]
\end{equation*}
 The Polyakov loop potential $\cal {U^\prime}$ with the Vandermonde
(VdM) term can be expressed as \cite{rayvdm},
\begin{equation}
{{\cal {U^\prime}}(\Phi,\bar \Phi,T)/ {T^4}}=
 {{\cal U}(\Phi,\bar \Phi,T)/ {T^4}}-\kappa \ln[J(\Phi,{\bar \Phi})]
\label {uprime}
\end{equation}
where ${\cal U}(\Phi,\bar \Phi,T)$ is the Landau-Ginsburg type potential 
given by \cite{ratti1},
\begin{equation}
  \frac{{\cal U}(\Phi, \bar \Phi, T)}{T^4}=-\frac{{b_2}(T)}{ 2}
                 {\bar \Phi}\Phi-\frac{b_3}{6}(\Phi^3 + \bar \Phi^3)
                 +\frac{b_4}{4}{(\bar\Phi \Phi)}^2
\label{LG_pot}
\end{equation}
with,
\begin{equation}
     {b_2}(T)=a_0+{a_1}(\frac{{T_0}}{T})+{a_2}(\frac{{T_0}}{T})^2+
              {a_3}(\frac{{T_0}}{T})^3, 
\label{b2}
\end{equation}
 and $b_3$, $b_4$ are constants. $T_0$ is the deconfinement temperature
in a pure gauge theory. The VdM determinant in eqn. 
(\ref{uprime}) is given by \cite{rayvdm},
\begin {equation*}
J[\Phi, {\bar \Phi}]=(27/24{\pi^2})(1-6\Phi {\bar \Phi}+\nonumber\\
4(\Phi^3+{\bar \Phi}^3)-3{(\Phi {\bar \Phi})}^2)
\end{equation*}
Here $\kappa$ is a phenomenological constant which is determined 
by reproducing the pressure calculated on Lattice. 
For the Polyakov loop potential we choose the parameters which 
reproduce the Lattice data of pure gauge thermodynamics \cite{boyd}. 
According to pure SU(3) lattice gauge theory value of $T_0$ is found 
to be $270~ \rm MeV$. However we took $T_0$ as $190~ \rm MeV$ to get the 
crossover temperature ($T_c$) consistent with the LQCD data.
Various thermodynamic quantities like scaled pressure, entropy and 
energy density are reproduced extremely well in Polyakov loop model 
using the ansatz (\ref{LG_pot}) and (\ref{b2}) with parameters summarized 
below, 
\begin {align}
       a_0=6.75,    a_1=-1.95,  a_2=2.625,
      a_3=-7.44,   b_3=0.75,  b_4=7.5,  T_0=190 {\rm MeV}\nonumber
\end {align}

There have been some work \cite{megias1} where the Polyakov loop parameters 
have been obtained by fitting with the full LQCD results rather than 
with the pure gauge theory results as done here. However, since we have fitted 
the pressure obtained in our model with that obtained in the full LQCD to obtain 
the parameter $\kappa$, the effect of full QCD is incorporated. In this work the 
Polyakov loop is a global object. One can improve on it by taking in to the consideration 
of quantum and local corrections as done in \cite{megias2,megias3}. However, 
in this work, we are mainly looking at the trends of different observables rather 
than exact matching with the LQCD results. As more and more refined LQCD 
results are coming up we understand that exact quantitative status is going to 
change. Hence we do not incorporate those involved calculations in this 
work. 

\begin{table}
\begin{center}
\begin{tabular}{|c|c|c|c|c|c|c|c|c|c|c|}
\hline
Interaction &$ m_u $&$ m_s $&$ \Lambda $&$ g_S \Lambda^2 $&$ g_D \Lambda^5 $&$
g_1 \times 10^{-21} $&$ g_2 \times 10^{-22}$&$ \kappa $&$
T_C $ \\
 &$\rm (MeV)$& $\rm (MeV)$&$\rm (MeV)$&  &  &$ \rm (MeV^{-8})$ & $ \rm (MeV^{-8})$ &
  & $\rm (MeV)$ \\
\hline
                                                                                
$UEP $&$ 5.5 $&$ 134.758 $&$ 631.357 $&$ 3.664 $&$ 74.636
$&$ 0.0 $&$ 0.0 $&$ 0.13 $&$ 181 $\\

$BEP$ & $ 5.5 $&$ 183.468 $&$ 637.720 $&$ 2.914 $&$ 75.968
$&$ 2.193 $&$ -5.890 $&$ 0.06 $&$ 169 $\\
 
\hline
\end{tabular}
\caption{ Parameters and $T_C$ for UEP and BEP type Lagrangians. }
\label{table1}
\end{center}
                                                                                
\end{table}

\vskip 0.2in
{\subsection{Taylor expansion of pressure}}
The pressure of the strongly interacting matter can be written as,
\begin {equation}
P(T,\mu_B,\mu_Q,\mu_S)=-\Omega (T,\mu_B,\mu_Q,\mu_S),
\label{pres}
\end {equation}
where $T$ is the temperature, $\mu_B$ is the baryon (B) chemical potential, 
$\mu_Q$ is the charge (Q) chemical potential and $\mu_S$ is the 
strangeness (S) chemical potential. From the usual thermodynamic
relations the first derivative of pressure with respect to
quark chemical potential $\mu_q$ is the quark number density and
the second derivative corresponds to the QNS.

 Our first job is to minimize the thermodynamic potential numerically with
respect to the fields $\sigma_u$, $\sigma_d$, $\sigma_s$, $\Phi$ and 
$\bar \Phi$. Using these values of the fields we get the mean field
value for pressure using the equation (\ref{pres}).
The scaled pressure obtained in a given range of chemical potential 
at a particular temperature can be expressed in a Taylor series as,
\begin {equation}
\frac{p(T,\mu_B,\mu_Q,\mu_S)}{T^4}=\sum_{n=i+j+k}c_{i,j,k}^{B,Q,S}(T) 
           (\frac{\mu_B}{T})^i (\frac{\mu_Q}{T})^j (\frac{\mu_S}{ T})^k
\end{equation}
where,
\begin {equation}
c_{i,j,k}^{B,Q,S}(T)={\frac{1}{i! j! k!} 
\frac{\partial^i}{\partial (\frac{\mu_B}{T})^i} 
\frac{\partial^j}{\partial (\frac{\mu_Q}{T})^j} 
\frac{\partial^k {(P/T^4)}}{\partial (\frac{\mu_S}{T})^k}}\Big|_{\mu_{q,Q,S}=0}
\end{equation}
The flavor chemical potentials  $\mu_u$, $\mu_d$, $\mu_s$ are related to 
$\mu_B$, $\mu_Q$, $\mu_S$ by,
\begin {equation}
  \mu_u=\frac{1}{3}\mu_B+\frac{2}{3}\mu_Q,~~~ 
  \mu_d=\frac{1}{3}\mu_B-\frac{1}{3}\mu_Q,~~~
  \mu_s=\frac{1}{3}\mu_B-\frac{1}{3}\mu_Q-\mu_S
\label{mureln1}
\end {equation}
Here the odd terms vanish due to CP symmetry at vanishing chemical potential
and the correlation functions with $i+j+k$ even are nonzero. 
In this work we evaluate the correlation coefficients up to fourth order
which are generically given by;
\begin{equation}
c_{i,j}^{X,Y}=\dfrac{1}{i! j!}\dfrac{\partial^{i+j}\left(P/T^4\right)}
{{\partial\left({\frac{\mu_X}{T}}\right)^i}{\partial\left({\frac{\mu_Y}{T}}\right)^j}}
\end{equation}
where, X and Y each stands for B, Q and S with $X\neq Y$.
To extract the Taylor coefficients, first the pressure is obtained as 
a function of different combinations of chemical potentials for each 
value of T and fitted to a polynomial about zero chemical potential
using the gnu-plot fit program \cite{gnu}. Stability of the fit has 
been checked by varying the ranges of fit and simultaneously keeping 
the values of least squares to $10^{-10}$ or even less.
 
\vskip 0.3in

{\section{Results and Discussion}
\label{results}}
\begin{figure}[htb]
\subfigure[] 
{\includegraphics [scale=0.9] {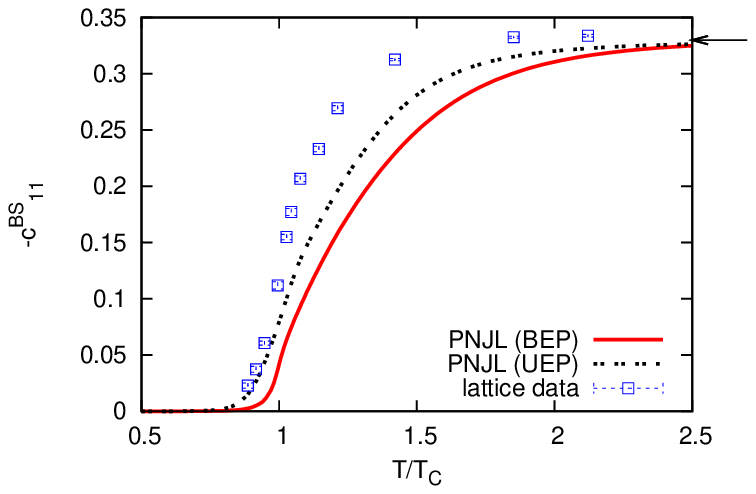}
\label {fg.bs11}}
\subfigure[]
{\includegraphics [scale=0.9] {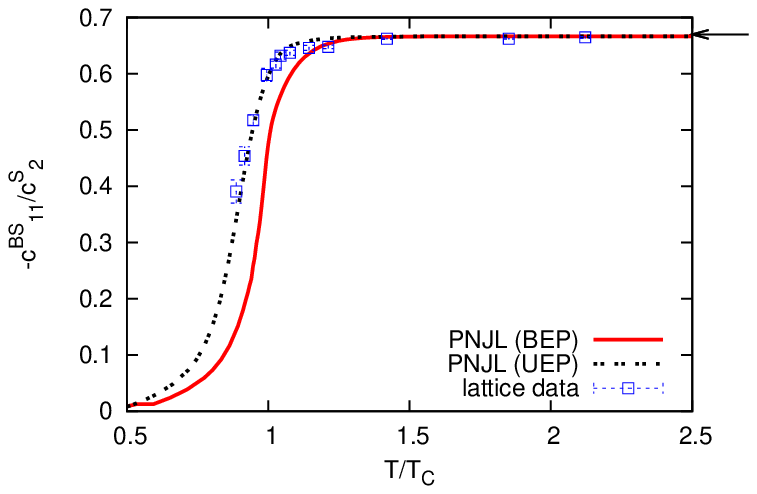}
\label {fg.bs11_s2}}
\subfigure[] 
{\includegraphics [scale=0.9] {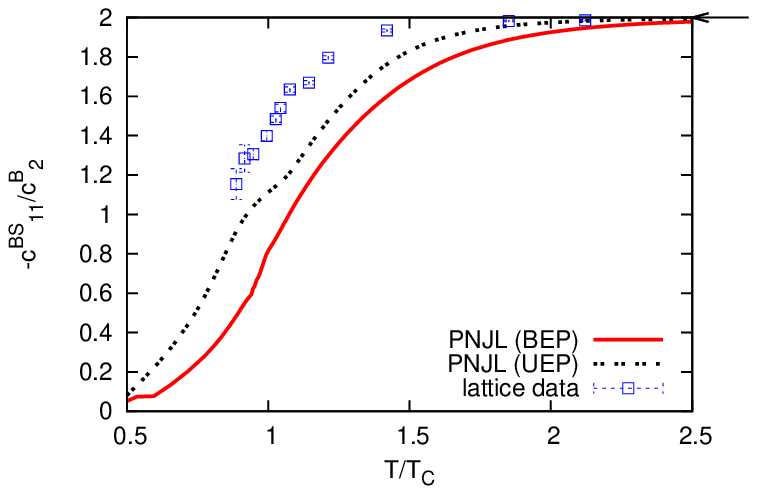}
\label {fg.bs11_b2}}
\vskip 0.1 in
   \caption{ Leading order Baryon-strangeness (BS)correlation as a function
of $T/T_C$. Lattice data taken from Ref.\cite{cheng}.
Arrows on the right indicate the corresponding SB limit.}
\label{fg.allbs2}
\end{figure}
We now set out to present the results obtained for correlation among
different conserved charges. First the leading order correlations 
are shown and compared with those of LQCD. Later we discuss the
behavior of some higher order correlations predicted from PNJL model.
\par
\begin{figure}[!t]
\subfigure[]
{\includegraphics [scale=0.9] {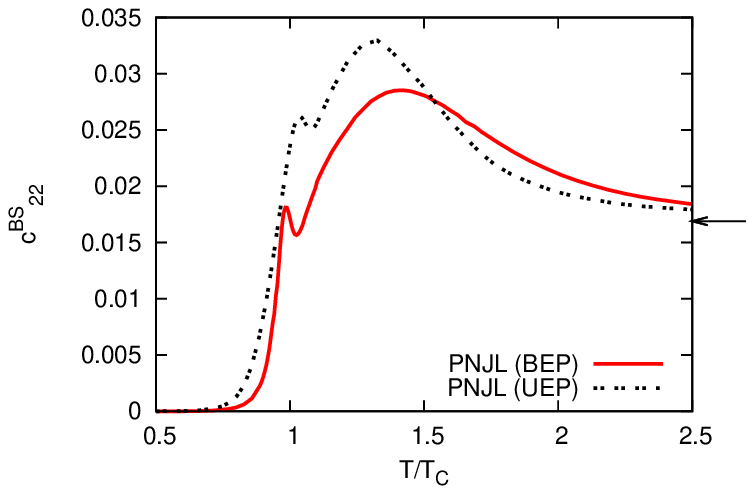}
\label {fg.bs22}}
\subfigure[]
{\includegraphics [scale=0.9] {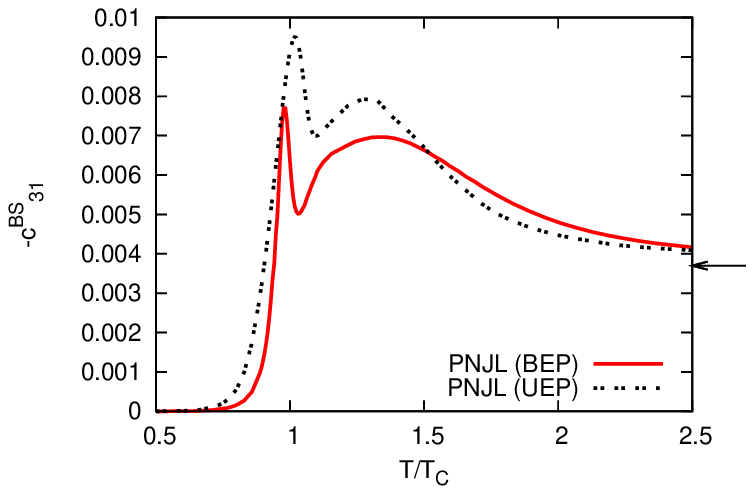}
\label {fg.bs31}}
\subfigure[]
{\includegraphics [scale=0.9] {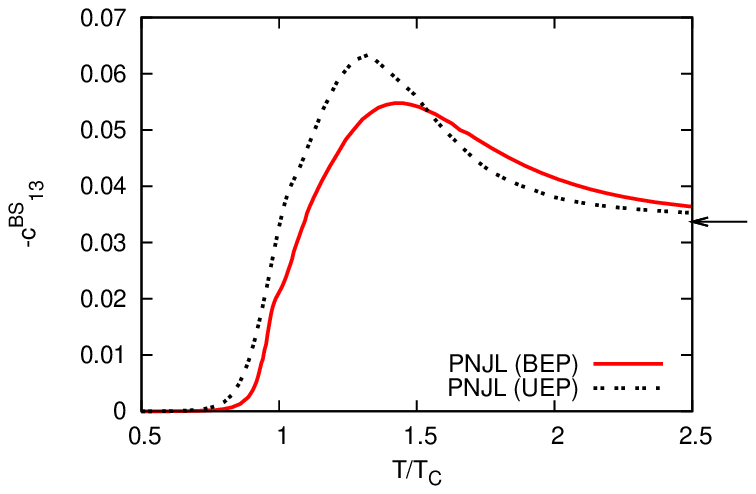}
\label {fg.bs13}}
\vskip 0.1 in
   \caption{ Fourth order baryon-strange correlation coefficients
as a function of $T/T_C$.
Arrows on the right indicate the corresponding SB limit.}
\label{fg.allbs4}
\end{figure}
Let us consider the baryon-strangeness (BS) correlation.
In Fig.\ref{fg.bs11} leading order BS correlation is shown and compared with
LQCD data. Though there is good qualitative agreement the results have
significant difference quantitatively. This is not very surprising
as the inherent physical masses of the constituents in our model and those
on the Lattice are substantially different - the ratio of pion to kaon mass
is about a factor of 2 larger on the Lattice than the physical value.
As we shall see this kind of departure remains for almost all the 
correlations measured on the Lattice and in PNJL model.

 The BS correlation normalized to the strangeness and baryon number
fluctuations respectively are given by
\begin{eqnarray*}
\textrm{C}_\textrm{BS} = - \dfrac{\chi_{BS}}{\chi_{SS}} = -\frac{1}{2} \dfrac{c_{11}^{BS}}{c_2^S} \\
\textrm{C}_\textrm{SB} = - \dfrac{\chi_{BS}}{\chi_{BB}} = -\frac{1}{2} \dfrac{c_{11}^{BS}}{c_2^B}
\end{eqnarray*}
where we have used the notation; 
$\chi_{XY}=\dfrac{\partial^2P}{\partial\mu_X\partial\mu_Y}$ and
$\chi_{XX}=\dfrac{\partial^2P}{\partial\mu_X^2}$.
It was argued in Ref.\cite{koch:2005} that $\textrm{C}_\textrm{BS}$ has 
entirely different behavior in hadron gas and in QGP, and therefore
this can be a reasonable diagnostic tool for identifying the
nature of the matter formed in heavy-ion collisions through
event-by-event fluctuations. In quark phase, baryon number and 
strangeness are strongly correlated through the strange quark indicating
$\textrm{C}_\textrm{BS}$ should approach its Stefan-Boltzmann (SB) limit
as soon as quark quasi-particles are dominant. The corresponding
value for the ratio $\dfrac{c_{11}^{BS}}{c_2^S}$ is -$\frac{2}{3}$. 
From Fig.\ref{fg.bs11_s2} we see that closely above $T_C$ the ratio 
reaches its SB limit.
\par
\begin{figure}[!t]
\subfigure[] 
{\includegraphics [scale=0.9] {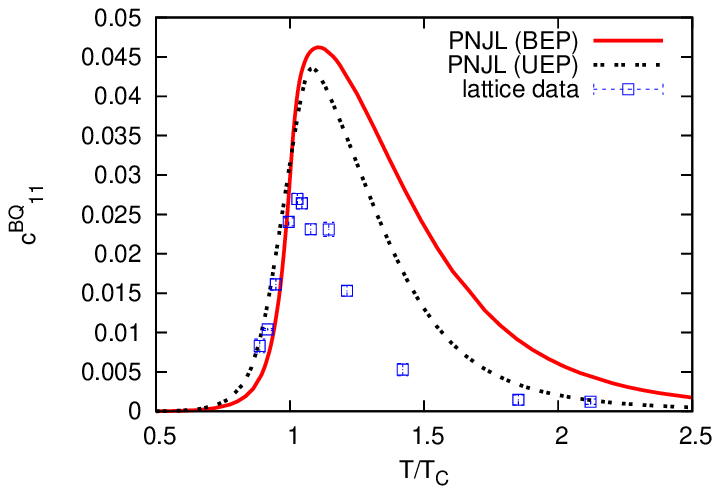}
\label {fg.bc11}}
\subfigure[] 
{\includegraphics [scale=0.9] {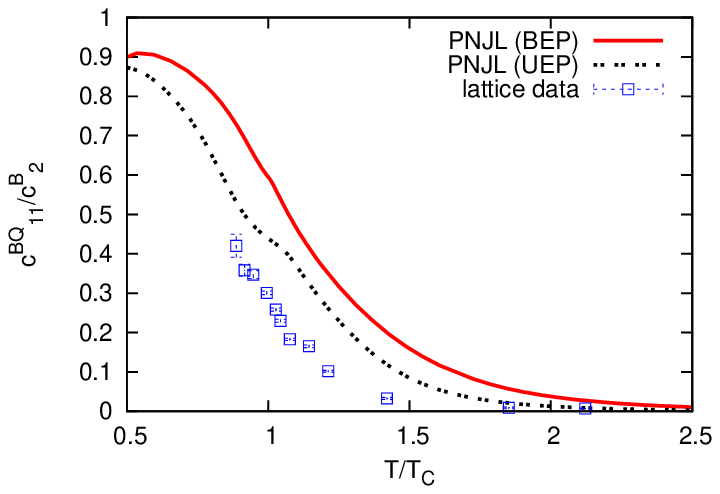}
\label {fg.bc11_b2}}
\subfigure[]
{\includegraphics [scale=0.9] {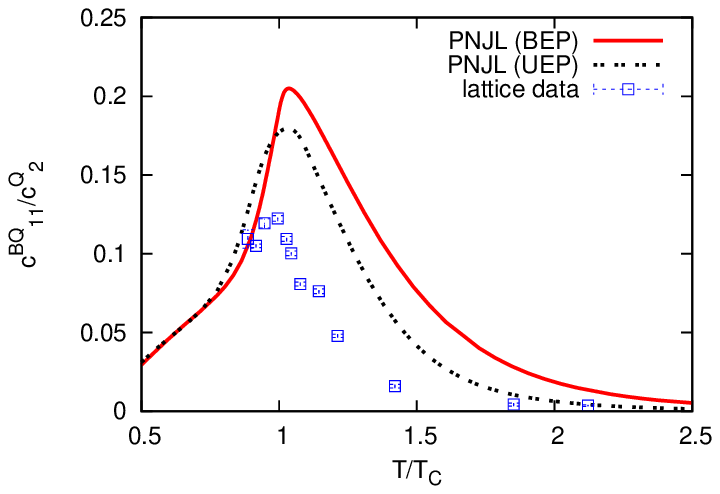}
\label {fg.bc11_c2}}
\vskip 0.1 in
   \caption{ Leading order baryon-charge (BQ) correlation
as a function of $T/T_C$. Lattice data taken from Ref.\cite{cheng}.}
\label{fg.allbc2}
\end{figure}
At low temperatures in the hadronic phase the situation is different. 
Numerator of $\textrm{C}_\textrm{BS}$ has contributions from strange 
baryons only, whereas the denominator has contributions from all
strange hadrons. So the ratio approaches zero as the temperature is
decreased.
\par
A similar behavior is found in the ratio $\textrm{C}_\textrm{SB}$ 
which gives the BS correlation normalized to the fluctuation 
of baryon number. This is shown in Fig.\ref{fg.bs11_b2}. 
Here again the high temperature behavior is consistent with a quark
quasi-particle picture and at the low temperatures the strange
baryon correlation is much smaller than the baryon fluctuation due
to the large mass of the strange baryons.
For both these ratios we see a nice qualitative agreement with lattice
data though quantitative disagreement persist. 
\begin{figure}[!t]
\subfigure[]
{\includegraphics [scale=0.9] {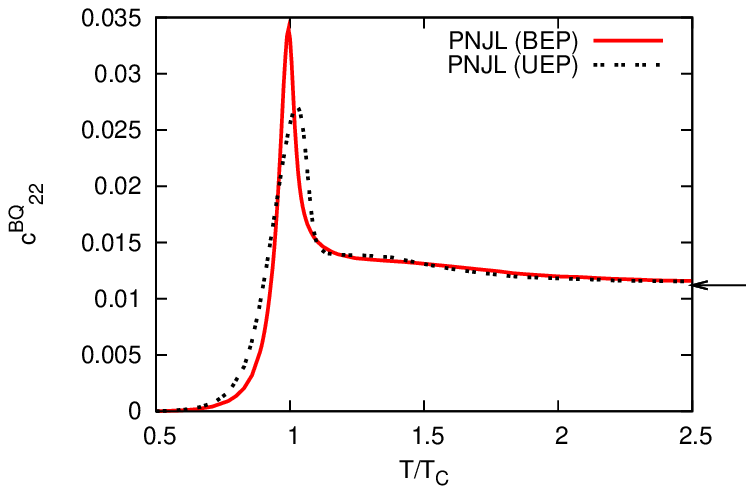}
\label {fg.bc22}}
\subfigure[]
{\includegraphics [scale=0.9] {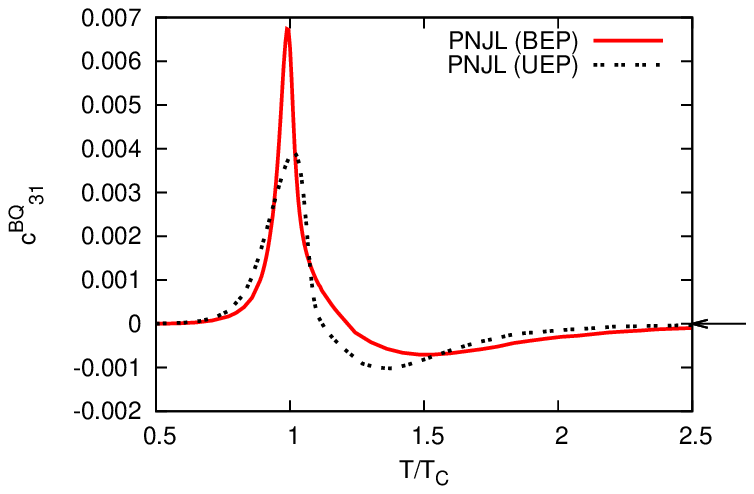}
\label {fg.bc31}}
\subfigure[]
{\includegraphics [scale=0.9] {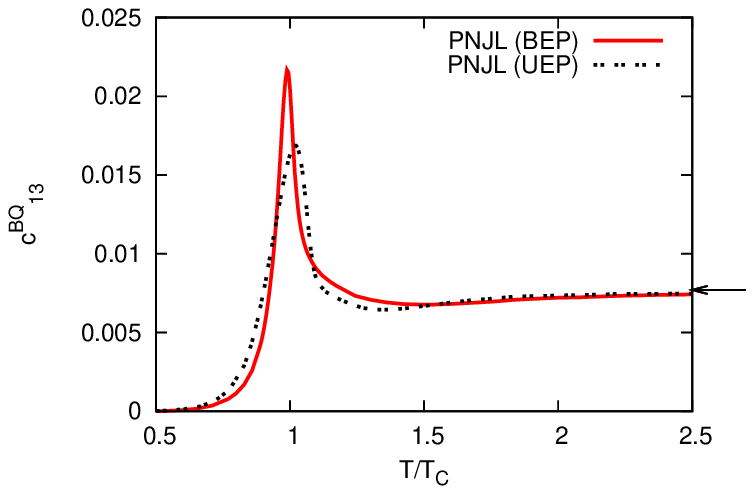}
\label {fg.bc13}}
\vskip 0.1 in
   \caption{ Fourth order baryon-charge correlation coefficients
as a function of $T/T_C$.
Arrows on the right indicate the corresponding SB limit.}
\label{fg.allbc4}
\end{figure}

\par
Going a step further we show the behavior of some fourth order correlations 
- $c^{BS}_{22}$, $c_{31}^{BS}$ and $c_{13}^{BS}$ in Fig.\ref{fg.allbs4}. 
At low $T$, in the hadronic phase, all three correlations go to zero. On the
other hand the correlations approach their SB limit at temperature 
close to 2.5$T_c$. For $c_{22}^{BS}$ and $c_{31}^{BS}$ there are two 
cusps, one corresponding to the chiral transition in the light quark
sector and the other corresponding to the strange sector. Similar features
for diagonal correlators were discussed by us in Ref. \cite{deb3}. For 
$c_{13}^{BS}$
the strange sector completely overwhelms the light quark sector as
expected, and so there is only one peak at 1.5$T_c$. Therefore, if
somehow these three correlations freeze out earlier than thermal and 
chemical freeze-out in heavy-ion collision experiments, they will not 
only be good indicators of the crossover but can also draw the explicit 
distinction between the chiral transitions in the light and strange 
quark sectors.
\begin{figure}[!htb]
\subfigure[] 
{\includegraphics [scale=0.9] {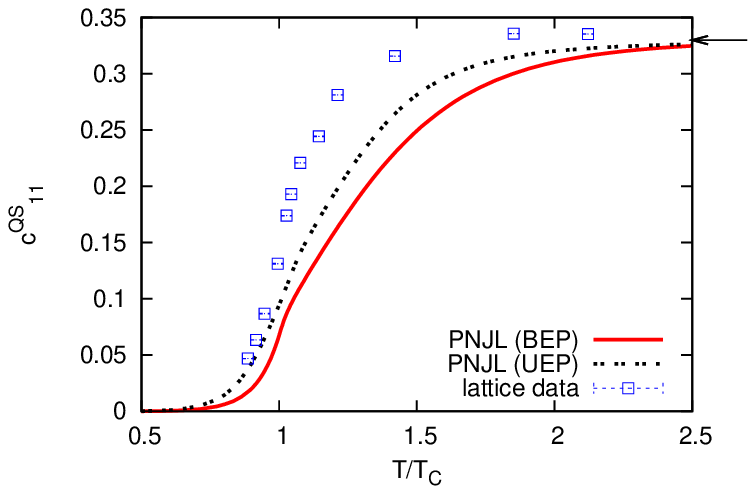}
\label {fg.cs11}}
\subfigure[]
{\includegraphics [scale=0.9] {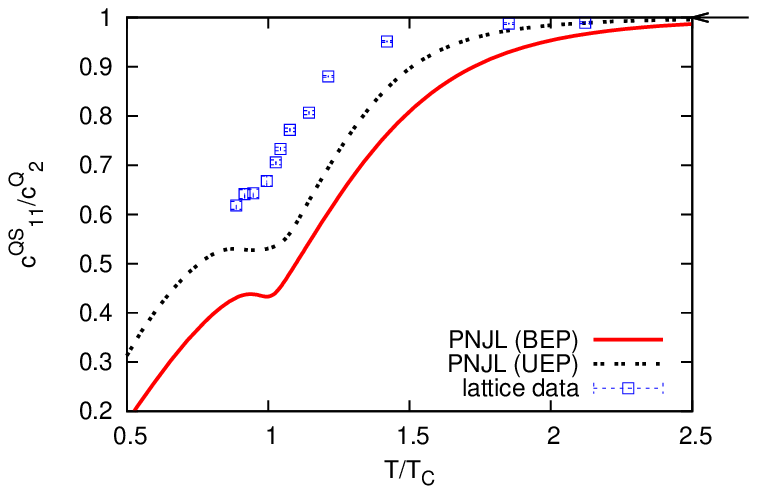}
\label {fg.cs11_c2}}
\subfigure[] 
{\includegraphics [scale=0.9] {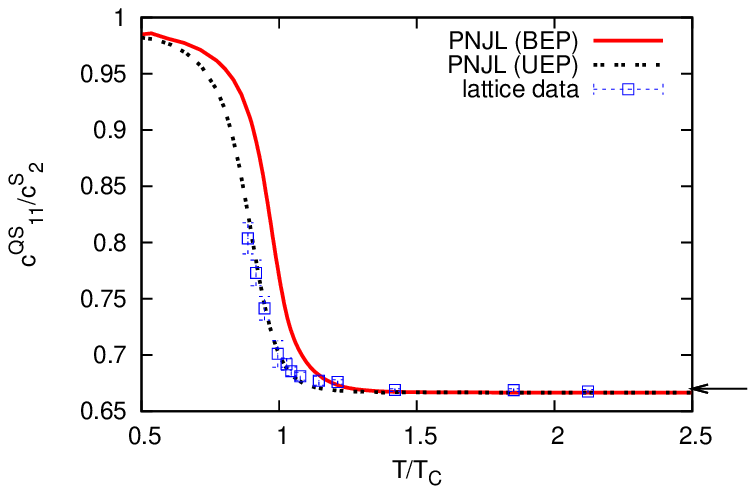}
\label {fg.cs11_s2}}
\vskip 0.1 in
   \caption{ Leading order charge-strange correlation 
as a function of $T/T_C$. Lattice data taken from
Ref.\cite{cheng}.
Arrows on the right indicate the corresponding SB limit.}
\label{fg.allcs2}
\end{figure}
\par
We now turn to baryon-charge (BQ) correlation. In Fig.\ref{fg.bc11} leading
order correlation is shown. At both very low and at very high temperatures
$c_{11}^{BQ}$ is zero. This is because at low temperatures the contributions
from heavy baryons decrease. On the other hand in the high temperature
weakly interacting phase the baryon and charge quantum numbers are
completely independent of each other. The peak in $c_{11}^{BQ}$ occurs
slightly above $T_c$ and is a clean indicator of the crossover. 
We show $c_{11}^{BQ}$ normalized with respect to $c_2^B$ and $c_2^Q$ 
respectively in other two panels of Fig.\ref{fg.allbc2}. These 
can be expressed in terms of the following:
\begin{eqnarray*} 
\textrm{C}_\textrm{BQ} &=& \dfrac{\chi_{BQ}}{\chi_{QQ}} = \frac{1}{2} \dfrac{c_{11}^{BQ}}{c_2^Q}\\
\textrm{C}_\textrm{QB} &=& \dfrac{\chi_{BQ}}{\chi_{BB}} = \frac{1}{2} \dfrac{c_{11}^{BQ}}{c_2^B}
\end{eqnarray*} 
Both the ratios go to zero at high $T$. At low $T$ the ratios show 
different behavior. While $c_2^B$ becomes small at low temperatures due 
to heavy baryons, $c_2^Q$ is not so small due to the contributions from
light mesons. Thus $\textrm{C}_\textrm{QB}$ remains non-zero whereas
$\textrm{C}_\textrm{BQ}$ goes to zero. Interestingly, in case of
a complete thermal equilibrium, in the fireball created in heavy-ion 
collisions, these two quantities, if measured, would give valuable
insight into the quantitative aspects of the PNJL model at low 
temperatures in the hadronic phase. The difference in Lattice and
PNJL model studies which is more pronounced below $T_c$, would then 
have to confront this measurement in the experiment.
\par
In Fig.\ref{fg.allbc4} we have plotted the 3 fourth order correlation
coefficients $c_{22}^{BQ}$, $c_{31}^{BQ}$ and $c_{13}^{BQ}$. All of them
show a pronounced peak close to $T_c$. Among these $c_{31}^{BQ}$ is
non-zero only at $T_c$ and so is the best indicator of the crossover.
These correlators do not show the double peak structure as contribution
from strange sector is subdominant.
\begin{figure}[t]
\subfigure[]
{\includegraphics [scale=0.9] {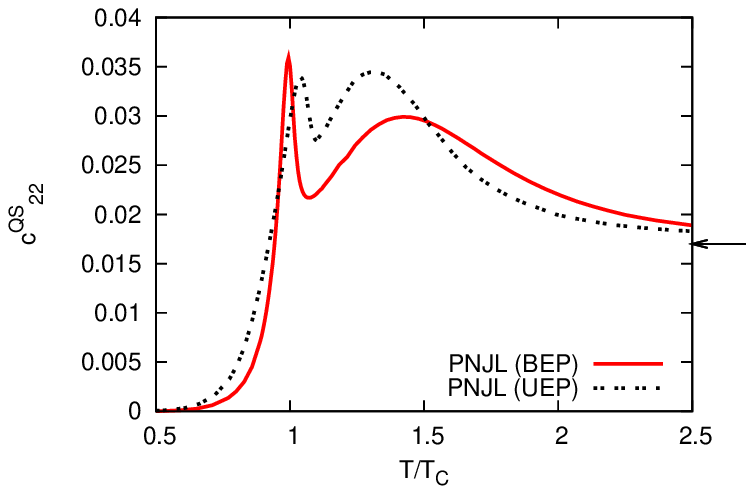}
\label {fg.cs22}}
\subfigure[]
{\includegraphics [scale=0.9] {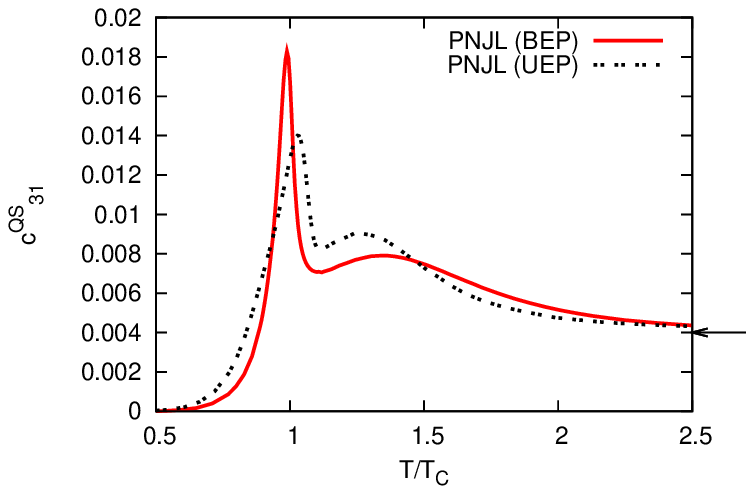}
\label {fg.cs31}}
\subfigure[]
{\includegraphics [scale=0.9] {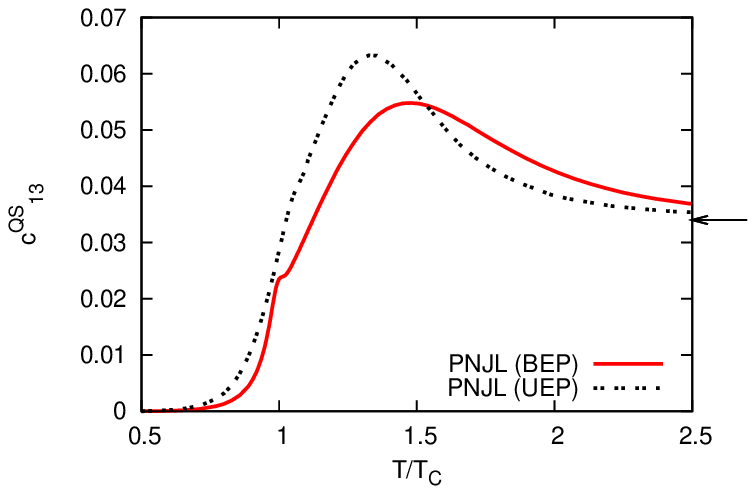}
\label {fg.cs13}}
\vskip 0.1 in
   \caption{ Fourth order charge-strange correlation coefficients
as a function of $T/T_C$.
Arrows on the right indicate the corresponding SB limit.}
\label{fg.allcs4}
\end{figure}
\par
In Fig.\ref{fg.allcs2} the leading order QS correlation is shown. As
in the case of baryon number, charge is also strongly correlated to
strangeness through strange quarks and therefore at high temperature
$c^{QS}_{11}$ approaches its SB limit. At low temperatures, the strangeness
carriers become heavier and hence the correlation decreases. Interesting
features arise in the ratios,
\begin{eqnarray*} 
\textrm{C}_\textrm{QS} &=& \dfrac{\chi_{QS}}{\chi_{SS}} = \frac{1}{2} \dfrac{c_{11}^{QS}}{c_2^S}\\
\textrm{C}_\textrm{SQ} &=& \dfrac{\chi_{QS}}{\chi_{QQ}} = \frac{1}{2} \dfrac{c_{11}^{QS}}{c_2^Q}
\end{eqnarray*}
The ratio $\textrm{C}_\textrm{SQ}$ decreases with decreasing temperature
until it reaches $T_c$. There it shows a plateau both in PNJL and Lattice
measurements as most likely this is where hadronic degrees of freedom
start to become dominant. Thereafter the ratio again decreases.
This seems to indicate that as we raise the temperature, the strangeness
content and hence its correlation keeps on increasing reaching a 
saturation. This is suddenly broken by the liberation of new degrees
of freedom in terms of quark quasi-particles just above $T_c$.
More detailed investigation is necessary to confirm this picture.
In retrospect we do expect and see a similar behavior for 
$\textrm{C}_\textrm{SB}$ and $\textrm{C}_\textrm{QB}$ though not highly 
prominent.

At large temperatures $\textrm{C}_\textrm{QS}$ takes up non-zero value
\cite{gavgup_fluc}. However it increases with
decreasing temperature due to the same reason as $\textrm{C}_\textrm{QB}$,
i.e., the strangeness fluctuations decrease at a faster rate at lower
temperatures. The saturation effect near $T_c$ is normalized out.

\par
In Fig.\ref{fg.allcs4} we have plotted the fourth order correlations. 
All three fourth order correlations $c_{22}^{QS}$, $c_{31}^{QS}$ and 
$c_{13}^{QS}$ show similar behavior as for the BS correlations.
Therefore these two sets can be used complementarily to understand
the state of affairs in heavy ion collisions.

\vskip 0.3in
{\section{Conclusion}}
We have studied the correlations between different conserved charges 
in the PNJL model. The baryon-strange (BS), baryon-charge (BQ) and the 
charge-strange (QS) correlations were obtained by fitting the pressure 
in a Taylor series expansion around vanishing chemical potential. 
The ratio of these leading order correlators to the respective quadratic
diagonal correlators were obtained. The results were shown both for
the physically bound effective PNJL model (BEP) as well as the conventional
PNJL model without a physical bound (UEP). Lattice QCD data exists for the
leading order correlators and were contrasted against our calculations.

In general the comparison with Lattice data gave excellent qualitative
agreement reproducing all the physical features in all the coefficients
that could be compared. However quantitative differences remain which
we believe is mainly due to the difference in effective masses. 
Interestingly the UEP seems to be closer to the Lattice results.
One can therefore say that the PNJL model is standing quite strong
as an effective model of QCD. 

As we discussed in the main text the combined studies of various 
correlators would give us a clearer picture of the matter created in
heavy-ion collision experiments. The leading order coefficients
can be most useful in identifying if the QGP is formed, while the
higher order coefficients could identify the crossover region.

We have noted a slight saturation of charge and strangeness through
the saturation of the ratios $\textrm{C}_\textrm{SQ}$, 
$\textrm{C}_\textrm{SB}$ and $\textrm{C}_\textrm{QB}$ in a small
temperature region just below $T_c$. This is found both in the PNJL
model and in LQCD. Perhaps this is when the degrees of freedom
crossover from hadronic to the partonic ones.

The higher order correlators containing strangeness show two cusps
at around $T_c$ and around 1.5$T_c$, corresponding to the temperatures
where the chiral crossover in the light and strange quark sectors occur.
This widens the temperature range where the fluctuations remain much
larger than the values at low temperature thus increasing the scope
of identifying the possible existence of the high temperature phase
in heavy-ion collisions. On the other hand the higher order BQ correlators
have a sharp peak and is an accurate tool to decide the crossover 
temperature.

Even if we are not lucky enough to identify the high temperature phase
from the correlations which need to get frozen much before thermal or
chemical freeze-out, the various equilibrium thermodynamic measurements
of the correlators would help us in determining the finite temperature
behavior of the hadronic sector studied theoretically using PNJL model 
and Lattice QCD.

\vskip 0.2 in
{\section{Acknowledgement}}
P.D. and A.L. would like to thank CSIR for financial support.
A.B. thanks CSIR and UGC (UPE and DRS) for support.
\vskip 0.2 in

{\it Note Added :} After finishing this work we learned of a recent work 
by W. Fu and Y. Wu (arxiv : 1010.0892 [hep-ph]). However they have 
calculated the off-diagonal susceptibilities only for the unbound 
effective potential of the PNJL model. Furthermore, they have not 
included the Vandermonde term in the Polyakov loop potential.


\begin{thebibliography}{99}

\bibitem{boyd} G. Boyd {\it{et al.}}, Nucl. Phys. B {\bf  469}, 419 (1996).

\bibitem{engels} J. Engels, O. Kaczmarek, F. Karsch, and E. Laermann, Nucl.
Phys. B {\bf  558}, 307 (1999).

\bibitem{fodor1} Z. Fodor and S. D. Katz, Phys. Lett. B {\bf  534}, 87 (2002).

\bibitem{fodor2} Z. Fodor, S. D. Katz and K. K. Szabo, 
Phys. Lett. B {\bf 568}, 73 (2003).

\bibitem{allton1} C. R. Allton {\it{et al.}}, Phys. Rev. D {\bf 66}, 
074507 (2002).

\bibitem{allton2} C. R. Allton {\it{et al.}}, Phys. Rev. D {\bf 68}, 
014507 (2003).

\bibitem{allton3} C. R. Allton {\it{et al.}}, 
Phys. Rev. D {\bf 71}, 054508 (2005).

\bibitem{forcrand} P. de Forcrand and O. Philipsen, Nucl. Phys. B 
{\bf 642}, 290 (2002); {\bf 673}, 170 (2003).

\bibitem{aoki1} Y. Aoki, Z. Fodor, S. D. Katz and K. K. Szabo, 
Phys. Lett. B {\bf 643}, 46 (2006).

\bibitem{aoki2} Y. Aoki, G. Endrodi, Z. Fodor, S. D. Katz and K. K. Szabo, 
Nature {\bf 443}, 675 (2006).

\bibitem{arriola1} E. Megias, E. Ruiz Arriola and L. L. Salcedo,
Rom. Rep. Phys. {\bf 58}, 081, (2006)
                                                                                
\bibitem{arriola2} E. Megias, E. R. Arriola and L. L. Salcedo,
Pos JHW2005, 025, (2006).
                                                                                
\bibitem{arriola3} E. Megias, E. R. Arriola and L. L. Salcedo,
Nucl. Phys. Proc. Suppl. {\bf 186}, 256, (2009).
                                                                                
\bibitem{arriola4} E. Megias, E. R. Arriola and L. L. Salcedo,
Phys. Rev. D {\bf 81}, 096009, (2010).

\bibitem{fuku1} K. Fukushima, Phys. Lett. B {\bf 591}, 277 (2004).

\bibitem{ratti1} C. Ratti, M. A. Thaler and W. Weise,
 Phys. Rev. D  {\bf 73}, 014019 (2006).

\bibitem{pisarski} R. D. Pisarski, Phys. Rev. D {\bf 62}, 111501 (2000);
A. Dumitru and R. D. Pisarski, Phys. Lett. B {\bf 504}, 282 (2001); 
{\bf 525}, 95 (2002); Phys. Rev. D {\bf 66}, 096003 (2002).

\bibitem{fuku2} K. Fukushima, Phys. Rev. D {\bf 77}, 114028, (2008).

\bibitem{ratti2} H. Hansen, W. M. Alberico, A. Beraudo, A. Molinari, M. Nardi 
and C. Ratti, Phys. Rev. D {\bf 75}, 065004 (2007).

\bibitem{gatto} M. Ciminale, R. Gatto, N. D. Ippolito, G. Nardulli and
 M. Ruggieri, Phys. Rev. D {\bf 77}, 054023 (2008).

\bibitem{ghosh} S. K. Ghosh, T. K. Mukherjee, M. G. Mustafa and R. Ray, 
Phys. Rev. D {\bf 73}, 114007 (2006).

\bibitem{rayvdm} S. K. Ghosh, T. K. Mukherjee, M. G. Mustafa and R. Ray, 
Phys. Rev. D {\bf 77}, 094024 (2008).

\bibitem{deb1} P. Deb, A. Bhattacharyya, S. Datta and S. K. Ghosh, 
Phys. Rev. C {\bf 79}, 055208 (2009).

\bibitem{osipov1} A. A. Osipov, B. Hiller and J. da Providencia, 
Phys. Lett. B {\bf 634}, 48, (2006).

\bibitem{osipov2} A. A. Osipov, B. Hiller, V. Bernard and A. H. Blin, 
Ann. Phys. {\bf 321}, 2504, (2006).

\bibitem{osipov3} A. A. Osipov, B. Hiller, A . H. Blin and J. da Providencia,
Ann. Phys. {\bf 322}, 2021, (2007).

\bibitem{osipov4} B. Hiller, J. Moreira, A. A. Osipov and A. H. Blin, Phys.
Rev. D {\bf 81}, 116005, (2010).

\bibitem{kashiwa1} K. Kashiwa, H. Kouno, T. Sakagauchi, M. Matsuzaki and 
M. Yahiro, Phys. Lett. B {\bf 647}, 446, (2007).

\bibitem{kashiwa2} K. Kashiwa, H. Kouno, M. Matsuzaki and 
M. Yahiro, Phys. Lett. B {\bf 662}, 26, (2008).

\bibitem{deb2} A. Bhattacharyya, P. Deb, S. K. Ghosh and R. Ray, 
Phys. Rev. D {\bf 82}, 014021 (2010) 

\bibitem{gottlieb} S. A. Gottlieb {\it{et al.}}, Phys. Rev Lett. {\bf 59}, 2247 
(1987).

\bibitem{gavai} R. V. Gavai, S. Gupta and P. Majumdar, Phys. Rev. D {\bf 65},
054506 (2002).

\bibitem{bernard1} C. Bernard {\it{et al.}}, Phys. Rev. D {\bf 71}, 034504 
(2005).

\bibitem{bernard2} C. Bernard {\it{et al.}}, Phys. Rev. D {\bf 77}, 014503 
(2008).

\bibitem{ejiri} S. Ejiri, F. Karsch and K. Redlich, Phys. Lett. B {\bf 633},
275 (2006).


\bibitem{roessner1} S. Roessner, C. Ratti and W. Weise, Phys. Rev. D
{\bf 75}, 034007, (2007).
                                                                                
\bibitem{friman} C. Sasaki, B. Friman and K. Redlich, Phys. Rev. D
{\bf 75}, 074013, (2007).
                                                                                
\bibitem{ray2} S. Mukherjee, M. G. Mustafa and R. Ray, Phys. Rev. D {\bf 75},
094015 (2007).
                                                                               
\bibitem{roessner2} C. Ratti, S. Roessner and W. Weise, Phys. Lett. B
{\bf 649}, 57, (2007).

\bibitem{mustafa1} P. Chakraborty, M. G. Mustafa and M. H. Thoma,
Eur. Phys. J. C {\bf 23}, 591-596, (2002).

\bibitem{mustafa2} P. Chakraborty, M. G. Mustafa and M. H. Thoma,
Phys. Rev D {\bf 68}, 085012, (2003).

\bibitem{mustafa3} N. Haque and M. G. Mustafa, arxiv:1007.2076 [hep-ph].
                                                                              
\bibitem{fuku3} K. Fukushima, Phys. Rev. D {\bf 79}, 074015, (2009).

\bibitem{cheng1} M. Cheng {\it{et al.}}, Phys. Rev. D {\bf 77}, 014511, (2008).

\bibitem{cheng} M. Cheng {\it{et al.}}, Phys. Rev. D {\bf 79}, 074505, (2009).

\bibitem{fodor10} S. Borsanyi  {\it{et al.}}, JHEP {\bf 1011}, 077, (2010).

\bibitem{wu1} W. J. Fu, Y. X. Liu and Y. L. Wu, Phys. Rev. D 
{\bf 81}, 014028, (2010).

\bibitem{wu2} W. J. Fu and Y. L. Wu, arxiv:1008.3684v1 (hep-ph), (2010).

\bibitem{deb3} A. Bhattacharyya, P. Deb, A. Lahiri and R. Ray, 
Phys. Rev. D {\bf 82}, 114028 (2010).

\bibitem{schaefer1} B. J. Schaefer and J. Wambach, Phys. Rev. D
{\bf 75}, 085015, (2007).
                                                                                
\bibitem{schaefer2} B. J. Schaefer, J. M. Pawlowski and J. Wambach, 
Phys. Rev. D {\bf 76}, 074023, (2007).
                                                                                
\bibitem{schaefer3}  B. J. Schaefer, M. Wagner and J. Wambach, Phys. Rev. D
{\bf 81}, 074013, (2010).
                                                                                
\bibitem{schaefer4}  J. Wambach, B. J. Schaefer and M. Wagner, 
Acta. Phys. Polon. Supp. {\bf 3} 691, 2010.
                                                                                
\bibitem{skokov}V. Skokov, B. Friman, E. Nakano, K. Redlich and B. J. Schaefer,
Phys. Rev. D {\bf 82} 034029 (2010).

\bibitem{megias1} E. Megias, E. Ruiz Arriola and L. L. Salcedo,
Eur. Phys. Jr.  A {\bf 31}, 553, (2007).

\bibitem{megias2} E. Megias, E. Ruiz Arriola and L. L. Salcedo,
Phys. Rev. D {\bf 74}, 065005, (2006). 

\bibitem{megias3} E. Megias, E. Ruiz Arriola and L. L. Salcedo,
AIP Conf. Proc. {\bf 892}, 444, (2007). 

\bibitem{gnu}{http://www.gnuplot.info/.}

\bibitem{koch:2005} V. Koch, A. Majumder and J. Randrup,
Phys. Rev. Lett. {\bf 95}, 182301, (2005).

\bibitem{gavgup_fluc} R. V. Gavai and S. Gupta,
Phys. Rev. D {\bf 73}, 014004 (2006).


\end{thebibliography}
\end{document}